\documentclass[prc,showpacs,floatfix]{revtex4}
\usepackage{bm}
\usepackage{graphicx}
\begin{document}
\title{
{\bf Analysis of angular distributions in $\gamma N\to\pi^0\eta N$}}

\author{A. Fix\footnote{Permanent address: Laboratory of Mathematical Physics,
Tomsk Polytechnic University, 634050 Tomsk, Russia}%
, M. Ostrick,  and L. Tiator } %
\affiliation{Institut f\"ur Kernphysik, Johannes Gutenberg-Universit\"at Mainz, D-55099 Mainz, Germany}

\date{\today}

\begin{abstract}
Angular distributions in the final state of $\pi^0\eta$ photoproduction on nucleons are considered. As a formal base the familiar isobar model
is used in which the $\pi^0\eta N$ state is a product of the resonance decay into $\eta\Delta(1232)$ and $\pi S_{11}(1535)$ channels. One of the
principal assumptions used is that in the actual energy region the reaction is dominated by a single resonance state. The developed formalism
can serve as a tool for testing spin and parity of that resonance.
\end{abstract}


\maketitle

\section{Introduction}

\begin{figure}[htb]
\begin{center}
\resizebox{0.60\textwidth}{!}{%
\includegraphics{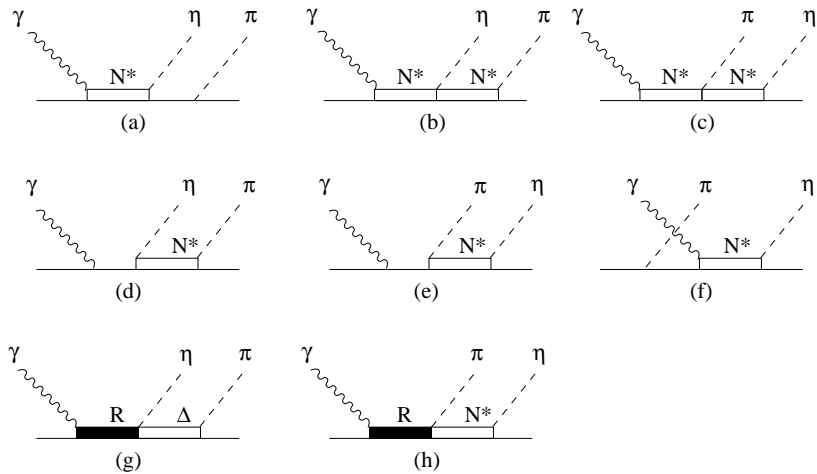}}
\caption{Diagrams for $\gamma N \to\pi^0\eta N$ used in the
calculations. The notations $\Delta$ and $N^*$ stand for $P_{33}(1232)$
and $S_{11}(1535)$, respectively.}
\label{fig1}
\end{center}
\end{figure}

Recent measurements~\cite{PiEtaDat} of the total cross section for $\pi^0\eta$ photoproduction off the proton
have shown that its value rapidly rises with increasing energy and in the region $E_\gamma>1.2$ GeV it exceeds the cross section for single
$\eta$ photoproduction. At the same time, already a simple analysis shows that the dynamics of this reaction is by no means trivial. Indeed,
calculations made with the Born diagrams, Fig.~\ref{fig1}(a-f), give only about 10$\%$ of the measured cross section value thus pointing to the
crucial role of resonances in this reaction (diagrams (g) and (h) in Fig.~\ref{fig1}).

The authors of Ref.~\cite{Doring} have proposed as a dominant mechanism excitation of the resonance $D_{33}(1700)$ with a subsequent decay into
the $\eta\Delta(1232)$ channel. It is quite clear that if the role of the Born terms is negligible, then in order to account for the rapid rise
of the cross section near threshold, one has no choice but to take a resonance decaying into $s$-wave $\eta\Delta$ state for which the
$D_{33}(1700)$ is a very good candidate.

However, \ according \ to \ the experimental results of Ref.\,\cite{PiEtaDat}, the cross section does not demonstrate a pronounced resonance
like energy dependence. It reaches about 4 $\mu$b at $E_\gamma=1.3$ GeV and then does not essentially change at least up to $E_\gamma=2.2$ GeV.
Such a smooth behavior might be due to the fact that the mass of the excited resonance is close or even below the threshold energy, so that the
corresponding peak is folded with the increasing reaction phase space. This situation might lead to strong uncertainties in determining the
resonance parameters when fitting the cross section. Furthermore we can not exclude the case with two or more strongly overlapping resonance
states, having different spin and parity and thus contributing incoherently. To clarify the situation a detailed analysis of partial waves
becomes of special importance. Some work in this direction has already been done by the Bonn-Gatchina group \cite{PiEtaDat}. Their method based
on the formalism of Ref.\,\cite{SaAn} has allowed them to analyse the role of different partial waves. According to their results in the region
$1.07$~GeV~$\leq E_\gamma\leq 1.45$~GeV the dominant contribution comes from the resonances $D_{33}(1700)$ and $D_{33}(1940)$.

The present study deals mainly with the angular dependence of the cross section which is proposed as a sensitive criterion to identify the
underlying production mechanism. Obviously, in the event that only one partial wave dominates, the corresponding angular distribution is
governed by the lowest powers of sine or cosine of spherical angles, whereas the Born sector, where large amount of partial waves are involved,
provide more or less monotonic angular dependence. Therefore, if we are able to separate the shape of the differential cross section associated
with pure harmonics we can expect that not only the resonance part of the amplitude can be isolated from the background but that it is also
possible to identify the quantum numbers of individual resonance states.

For simplicity as a first step we consider an ideal case when in the actual energy region of energy the amplitude is dominated by only one
resonance $R$ and contributions of other waves can be neglected. We put the resonance mass to $M_R=1.8$~GeV and take the total width
$\Gamma_R=300$ MeV. The quantum numbers, spin and parity $J^\pi$, are treated as model parameters. Our aim is to investigate the dependence of
the angular distributions on the choice of $R(J^\pi)$ and in this way to study the signature of individual partial waves. Perhaps, the next
natural move would be to identify $R(J^\pi)$ with more or less well-established resonance states known, e.g., from the PDG listing~\cite{PDG}.
However, within the quite rudimentary state of the database
and the theoretical descriptions, such an approach cannot lead to a unique solution.

The paper is organized as follows. First, in Sect. II we briefly describe the formalism of the sequential decay of the resonance $R$ according
to the scheme: $R\to$ baryon($\frac32^+$)$+$ meson($0^-$), baryon($\frac32^+$) $\to$ baryon($\frac12^+$) $+$ meson($0^-$) and derive expressions
for the angular distributions. Then in Sect.\,\ref{results} using the assumption of the isobar model we calculate the angular and energy
dependence
within different hypotheses about spin-parity of the resonance $R$. Finally, in Sect. IV we summarize our main qualitative results.

\section{Angular distributions. Formalism.}\label{model}

We consider the process
\begin{equation}\label{10}
\gamma(\vec{k}\,)+N_i(-\vec{k})\to\pi^0(\vec{q}_\pi)+\eta(\vec{q}_\eta)+N_f(\vec{p}_f\,)\,,
\end{equation}
where the 3-momenta of the particles in the overall c.m.\ frame are given in parenthesis. In general, the resonance mechanism of the reaction
(\ref{10}) can be realized according to the following two schemes (throughout the paper the resonance $S_{11}(1535)$ is denoted by $N^*$)
\begin{eqnarray}\label{15}
(a):\ \gamma N\to R(J^\pi)\to\eta\Delta^+\to \pi^0\eta N\,, \\
(b):\ \gamma N\to R(J^\pi)\to\pi^0 N^*\to \pi^0\eta N\, \nonumber
\end{eqnarray}
(see diagrams (g) and (h) in Fig.\,\ref{fig1}), the relative amount of which depends on the details of the reaction dynamics. The first scheme
with $R=D_{33}(1700)$ was considered in \cite{Doring} as a main driving mechanism of the reaction (\ref{10}). The second sequence appears in
Ref.~\cite{Doring} due to strong $\eta N$ interaction via $N^*$ excitation.

The resonance states $R$ considered in this paper are listed in Table~\ref{ta1} together with orbital momenta associated with their decay into
different channels. Throughout this section we assume that the $\pi^0\eta$ production always proceeds according to the scheme (a). The scheme
(b) will be included in the next section where we present our results obtained within the isobar model.

When writing the resonances in the form $L_{2T2J}$ in Table~\ref{ta1} we took into account that only the states with isospin $T=3/2$ decay into
the $\eta\Delta$ channel. As already noted, we do not try to identify the states $R(J^\pi)$ with the baryon spectrum known from PDG~\cite{PDG}.
But if only the quantum numbers are taken into consideration the states collected in Table~\ref{ta1} may be identified with, e.g.,
$S_{31}(1900)$, $P_{31}(1910)$, $D_{33}(1700)$, $P_{33}(1920)$, $D_{35}(1930)$, and $F_{35}(1905)$. These resonances, except for $D_{33}(1700)$,
belong to the third group and can influence the low-energy region only through their large widths. It is also worth noting that all the
mentioned states are characterized by quite a weak $\pi N$ mode (generally less than 20\,$\%$) and therefore can intensively decay into the
two-meson channels.

\begin{figure*}
\begin{center}
\resizebox{0.75\textwidth}{!}{%
\includegraphics{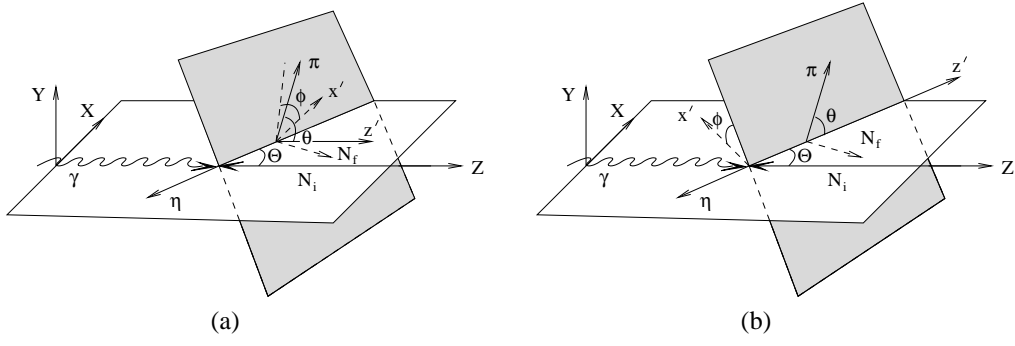}}
\caption{The angles $\Theta$ and $\Omega=(\theta,\phi)$ describing directions of the final particles. The panels (a) and (b) represent
configurations of the momenta in the canonical and the helicity systems, referred in the text as $K$- and $H$-system, respectively.}
\label{figXYZ}
\end{center}
\end{figure*}

The kinematics of the reaction (\ref{10}) is presented in Fig.\,\ref{figXYZ}. We select the $z$-axes along the photon momentum $\vec{k}$. The
production plane is spanned by the momenta $\vec{k}$ and $\vec{q}_\eta$, so that $\phi_\eta=0$. The decay plane of the $\pi N$ pair is fixed by
the momenta $\vec{q}_\pi$ and $\vec{p}_f$. We denote by $\Omega=(\theta,\phi)$ the solid pion angle in the $\pi N$ rest frame. To describe the
$\Delta\to\pi N$ decay, two types of the coordinate systems $Ox'y'z'$ are used. In the first one, the canonical frame (also referred to as Adair
frame, Fig.\,\ref{figXYZ}(a)), all three axes are codirectional to those of the $OXYZ$ system, so that the system $O'x'y'z'$ is deduced by the
Lorentz boost determined by the vector $\vec{q}_\pi+\vec{p}_f$. In the helicity frame (Fig.\,\ref{figXYZ}(b)) the $z'$ axis is aligned along the
vector $\vec{q}_\eta+\vec{p}$ and the $x'$ axis is in the production plane. To shorten the notations we denote by $\Theta$ the polar angle of
the vector $\vec{p}_\Delta=\vec{q}_\pi+\vec{p}=-\vec{q}_\eta$ in the overall c.m. system, and by $\Omega=(\theta,\phi)$ the pion momentum in the
$\pi N$ rest frame. The coordinate systems above will further be referred as $K$- and $H$-system respectively. Clearly, they are connected to
each other by a rotation with the angle $\Theta$ around the $y'$ axis.

\begin{table*}
\renewcommand{\arraystretch}{2.0}
\caption{Angular momenta associated with a decay of the resonance $R(J^\pi)$  into hadronic channels. In the $\eta\Delta$ states only the lower
of two possible values of $L$ is assumed. The resonance $S_{11}(1535)$ is denoted as $N^*$.} \label{ta1}
\begin{center}
\begin{tabular*}{12cm}
{@{\hspace{0.6cm}}c@{\hspace{0.6cm}}|@{\hspace{1.6cm}}c
@{\hspace{1.6cm}}c@{\hspace{1.6cm}}c}  \hline\noalign{\smallskip}
$J^\pi(L_{2T2J})$ & $L(\pi N)$ & $L(\eta\Delta)$ & $L(\pi N^*)$ \\
\noalign{\smallskip}\hline\noalign{\smallskip}
$\frac12^-(S_{31})$  & 0 & 2 & 1 \\
$\frac12^+(P_{31})$  & 1 & 1 & 0 \\
$\frac32^-(D_{33})$  & 2 & 0 & 1 \\
$\frac32^+(P_{33})$  & 1 & 1 & 2 \\
$\frac52^-(D_{35})$  & 2 & 2 & 3 \\
$\frac52^+(F_{35})$  & 3 & 1 & 2 \\
\noalign{\smallskip}\hline
\end{tabular*}
\end{center}
\end{table*}

The amplitude associated with the diagram (g) in Fig.~\ref{fig1} has the following form in the canonical and the helicity frame
\begin{equation}\label{20}
t^K_{m_f\lambda}(\Theta,\Omega)= \frac{1}{\sqrt{4\pi}}\sum_{R(JL)}\alpha^RA^R_\lambda\sum_{m}\sqrt{2L+1} C^{\frac32M_\Delta}_{1m\,\frac12m_f}
C^{J\lambda}_{LM_L\,\frac32M_\Delta}\,Y^*_{1m}(\Omega)\,d^L_{M_L0}(\Theta)\,,
\end{equation}
\begin{equation}\label{20a}
t^H_{m_f\lambda}(\Theta,\Omega)=\frac{1}{\sqrt{4\pi}}\sum_{R(JL)}\alpha^RA^R_{\lambda} \sum_{m}\sqrt{2L+1}
C^{\frac32\mu_\Delta}_{1m\frac12m_f}C^{J\mu_\Delta}_{L0\,\frac32\mu_\Delta}\, Y^*_{1m}(\Omega)\,d^J_{\lambda\mu_\Delta}(\Theta) \,.
\end{equation}
Here, $C^{JM}_{j_1m_1\,j_2m_2}$ are the usual Clebsch-Gordan coefficients for the coupling $\vec{j}_1+\vec{j}_2=\vec{J}$. The index $L$ is the
$\eta+\Delta$ orbital momentum and $d^L_{mm'}(d^J_{\mu\mu'})$ are the rotation matrices. In the second equation $\mu_\Delta$ stands for the
$\Delta$ helicity. The spherical function $Y_{1m}(\Omega)$ specifies the angular dependence of the decay $\Delta\to\pi N$ in the $\pi N$ rest
frame. The index $\lambda$ is the initial state helicity. Since the momenta of the initial particles are along the $Z$-axes, the photon helicity
is combined with the nucleon helicity to give the $z$ projection of the total angular momentum $J$ equal to $\lambda\in\{\pm 1/2,\pm 3/2\}$. The
corresponding amplitude of the transition $\gamma N\to R(J^\pi)$ is denoted by $A^R_\lambda$. Parity conservation requires that
\begin{equation}\label{20b}
A^R_{-\lambda}=(-1)^{\frac12+J-l}A^R_{\lambda}\,.
\end{equation}
The parameters $\alpha^R$ characterizing the individual resonances $R$ contain constants and energy dependent functions (resonance propagators,
barrier penetration factors, coupling constants, etc.), which detailed structure is irrelevant for further discussions. We note that the
nucleon magnetic quantum numbers $m_f$ in (\ref{20}) and (\ref{20a}) are projections of the final nucleon spin on different $z'$-axes, according
to their definitions in the $K$- and $H$-system (see Fig.\,\ref{figXYZ}).

Using Eq.\,(\ref{20}) or (\ref{20a}) together with (\ref{20b}) it is easy to verify that parity conservation leads to the following symmetry
property
\begin{equation}\label{25}
t^{K/H}_{-m_f-\lambda}(\Theta;\theta,\phi)=(-1)^{m_f-\lambda}\,
t^{K/H}_{m_f\lambda}(\Theta;\theta,-\phi)\,.
\end{equation}
The square of the matrix element can easily be written down from (\ref{20}) and (\ref{20a}). Since in this section our main object is the form
of angular distributions and not the absolute value of the cross section, it is convenient to introduce the distribution functions, normalized
to unity. In the isobar model discussed in the next section we assume that the only nonvanishing contribution to the amplitude comes from the
sole state $R(J^\pi)$, whereas other resonances can be neglected. Deriving the formulas below we will always adhere this somewhat oversimplified
picture.

First, we consider the distribution over the angle $\Theta$ of the $\pi N$ system in the overall c.m.\ frame. For this purpose we define the
distribution function
\begin{equation}\label{30}
W(\Theta)=\frac{\pi}{N}\int \sum\limits_{m_f\lambda}\left|\, t_{m_f\lambda}(\Theta;\Omega)\right|^2\,d\Omega\,,
\end{equation}
\begin{equation}\label{30a}
N=\left|\,\alpha^R\right|^2\Big(\left|\,A^R_{1/2}\right|^2+\left|\, A^R_{3/2}\right|^2\Big)\,,\quad \int\limits_0^\pi W(\Theta)\sin\!\Theta\,
d\Theta=1\,.\nonumber
\end{equation}
Using (\ref{20}) for a given resonance $R(J^\pi)$ we obtain
\begin{equation}\label{35}
W(\Theta)=\frac{2L+1}{4(1+a)}\sum_{M_L}|\, d^L_{M_L0}(\Theta)|^2\Big(\sum_{\lambda=\pm\frac12}(C^{J\lambda}_{LM_L\, \frac32\lambda-M_L})^2
+a\!\sum_{\lambda=\pm\frac32}(C^{J\lambda}_{LM_L\, \frac32\lambda-M_L})^2\Big)\,,
\end{equation}
where the parameter $a$ is defined as
\begin{equation}\label{35a}
a=\left(\frac{A^R_{3/2}}{A^R_{1/2}}\right)^2\,.
\end{equation}
As is evident from Eq.~(\ref{35}), the distribution over $\cos\Theta$ is flat for $J^\pi=1/2^\pm$ since in this case $a=0$ and \\
$\sum\limits_{\lambda=\pm 1/2}(C^{\frac12\lambda}_{LM_L\,\frac32\lambda-M_L})^2=2/(2L+1)$ does not depend on $M_L$. The same is true for $L=0$
which we have in the state with $J^\pi=3/2^-$ (see Table~\ref{ta1}). Thus
\begin{equation}\label{40}
W(\Theta)=\frac{1}{2}\,, \quad \mbox{if}\ \ J^\pi=\frac12^\pm\ \
\mbox{or}\quad J^\pi=\frac32^-(L=0)\,.
\end{equation}
Otherwise the shape of $W(\Theta)$ is described by a polynomial of $\cos\Theta$ of the order $2L$. Table~\ref{ta5} lists the function
$W(\Theta)$ for all six transitions considered here. As is seen from the Eq.~(\ref{35}), the exact form of $W(\Theta)$ depends on the ratio
$a=(A^R_{3/2}/A^R_{1/2})^2$. This fact brings a model dependence into our analysis, especially if the electromagnetic amplitudes $A_\lambda^R$
are poorly known. On the other hand, for $J^\pi=5/2^+$ all coefficients in the expansion are positively defined ($a>0$). As a result, the
corresponding distribution will always reach its maximum at $|\cos\Theta\,|=1$. For $J^\pi=3/2^+$ and $J^\pi=5/2^-$ the shape of the
distribution might be quite sensitive to the value of $a$. In such a situation indirect information about the spin of $R$ can be obtained from
the complexity of the angular distribution, which as mentioned before is fixed to $2L$ for $J\geq\frac32$.

\begin{table*}
\renewcommand{\arraystretch}{2.0}
\caption{Distribution $W(\Theta)$ (\protect{\ref{30}}) over the angle $\Theta$ of the
$\pi N$ system in the overall c.m.\ frame and the $\theta$
distribution $f(\theta)=2\pi\,W(\Theta=0;\theta,\phi)$~(\protect{\ref{80}}) of
pions produced by the $\Delta$ decay in coincidence with $\eta$
mesons at $\Theta_\eta=\pi$.} \label{ta5}
\begin{center}
\begin{tabular*}{\textwidth}{@{\hspace{0.4cm}}c@{\hspace{0.4cm}}|@{\hspace{0.5cm}}c
@{\hspace{0.5cm}}|@{\hspace{0.9cm}}c@{\hspace{0.3cm}}} \hline\noalign{\smallskip}
    $J^\pi(L_{2T2J})$ & $W(\Theta)$ & $f(\theta)$ \\
\noalign{\smallskip}\hline\noalign{\smallskip}
$\frac12^-(S_{31})$ & $\frac12$ & $\frac14(1+3\cos^2\theta)$ \\
$\frac12^+(P_{31})$ & $\frac12$ & $\frac14(1+3\cos^2\theta)$ \\
$\frac32^-(D_{33})$ & $\frac12$ & $\frac{1}{4(1+a)}(1+3a+3(1-a)\cos^2\theta)$\\
$\frac32^+(P_{33})$ & $\frac{1}{10(1+a)}(7+3a-6(1-a)\cos^2\Theta)$ &
$\frac{1}{4(1+9a)}(1+27a+3(1-9a)\cos^2\theta)$ \\
$\frac52^-(D_{35})$ & $\frac{3}{28(1+a)}(2+7a+5(4-5a)\cos^2\Theta-10(2-3a)\cos^4\Theta)$ &
$\frac{1}{4(1+6a)}(1+18a+3(1-6a)\cos^2\theta)$ \\
$\frac52^+(F_{35})$ & $\frac{3}{20(1+a)}(2+3a+(4+a)\cos^2\Theta)$ &
$\frac{3}{4(3+2a)}(1+2a+(3-2a)\cos^2\theta)$ \\
\noalign{\smallskip}\hline
\end{tabular*}
\end{center}
\end{table*}

Now we turn to the $\Omega$ dependence at fixed $\Theta$. The corresponding distribution function reads
\begin{equation}\label{45}
\displaystyle W(\Theta;\Omega)=\frac{1}{N(\Theta)}\,
\sum\limits_{m_f\lambda}\left|\,t_{m_f\lambda}(\Theta;\Omega)\right|^2\,,
\end{equation}
where the factor $N(\Theta)$ is determined by the normalization condition
\begin{equation}\label{45a}
\int W(\Theta;\Omega)\,d\Omega=1\,.
\end{equation}
It is convenient to present the function $W(\Theta;\Omega)$ in the form
\begin{equation}\label{50}
W(\Theta;\Omega)=\sum\limits_{mm'}Y^*_{1m}(\Omega)\,\rho_{mm'}(\Theta)Y_{1m'}(\Omega)\,,
\end{equation}
where, e.g., from Eq.\,(\ref{20}) the correlation coefficients $\rho_{mm'}(\Theta)$ read
\begin{equation}\label{55}
\rho_{mm'}(\Theta)=\frac{2L+1}{4\pi N(\Theta)}\sum_{m_f\lambda} \sum_{M_LM_L'}{\cal B}^{M_LM_L'}_{m_f\lambda,mm'} d^L_{M_L0}(\Theta)\
d^{L}_{M_L'0}(\Theta)
\end{equation}
with
\begin{equation}\label{60}
{\cal B}^{M_LM_L'}_{m_f\lambda,mm'}=\sum_{m_f\lambda}C^{\frac32M_\Delta}_{1m\, \frac12m_f}C^{\frac32M_\Delta'}_{1m'\,\frac12m_f}
C^{J\lambda}_{LM_L\,\frac32M_\Delta} C^{J\lambda}_{LM_L'\frac32M_\Delta'}\left|\,\alpha^RA^{R}_\lambda\right|^2.
\end{equation}
The unit-trace condition for the matrix $\rho_{mm'}$ immediately
follows from the Eqs.\,(\ref{45a}) and (\ref{50})
\begin{equation}\label{65}
\sum_m\rho_{mm}(\Theta)=1\,.
\end{equation}
It is intuitively clear that the structure of the pion angular distribution will be  governed by the $\Delta$ spin $J_\Delta=3/2$ and should
contain polynomials of $\cos\theta$ up to the order $2J_\Delta-1=2$. Using Eq.\,(\ref{50}) one has
\begin{eqnarray}\label{70}
W(\Theta;\theta,\phi)&=&\frac{3}{4\pi}\, \Big(\rho_{00}\cos^2\theta+\frac12(\rho_{11}+\rho_{-1-1})\sin^2\theta
-\sin^2\theta \left({\cal R}e\rho_{1-1}\cos2\phi-{\cal I}m\rho_{1-1}\sin2\phi\right)\nonumber\\
&+&\frac{1}{\sqrt{2}}\sin2\theta\,{\cal R}e(\rho_{-10}-\rho_{10})\cos\phi -\frac{1}{\sqrt{2}}\sin2\theta\, {\cal I}m(\rho_{-10}+\rho_{10})
\sin\phi\Big)\,,
\end{eqnarray}
where we have dropped the argument $\Theta$ in $\rho_{mm'}(\Theta)$. In (\ref{70}) the hermiticity of the matrix $\rho_{mm'}$ was already used.
Furthermore from (\ref{25}) it is evident that all elements $\rho_{mm'}$ are real and $\rho_{-m-m'}=(-1)^{m-m'}\rho_{mm'}$. Taking also into
account the normalization condition (\ref{65}) we arrive at the result that from nine real elements $\rho_{mm'}$ only three, $\rho_{00}$,
$\rho_{10}$ and $\rho_{1-1}$ remain independent so that the distribution function is reduced to
\begin{equation}\label{75}
W(\Theta;\theta,\phi)=\frac{3}{4\pi}\,\Big(\rho_{00}\cos^2\theta+\frac12(1-\rho_{00})\sin^2\theta -\sqrt{2}{\cal
R}e\rho_{10}\,\sin2\theta\cos2\phi-\rho_{1-1}\,\sin^2\theta\cos2\phi\Big)\,.
\end{equation}
Projection of (\ref{75}) on the $y'$ axis gives
\begin{equation}\label{70a}
W(\Theta;\theta=\phi=\frac{\pi}{2})=\frac{3}{4\pi}\,\Big(\frac12(1-\rho_{00})+\rho_{1-1}\Big)\,.
\end{equation}
Obviously, this combination is invariant under rotation around the $y'$ axis.  In fact, it is proportional to one of three eigenvalues of the
matrix $\rho$ (see, e.g., \cite{Don}) so that
\begin{equation}\label{70b}
\rho^H_{00}-2\rho_{1-1}^H=\rho^K_{00}-2\rho_{1-1}^K\,,
\end{equation}
where $\rho^H_{mm'}$ and $\rho^K_{mm'}$ are the correlation coefficients calculated in the $H$- and $K$-system.

Equation (\ref{75}) is the basic equation that will be used in the rest of the paper to evaluate the angular distributions. Its structure is
independent of the particular frame chosen for the description of the $\Delta$ decay, since it is fixed  by the $\Delta$ spin and the parity
conservation condition (\ref{25}).  The coefficients $\rho_{mm'}(\Theta)$ are determined by the spin-parity of the resonance $R$. Therefore,
analysis of the experimental angular dependence should enable one to get information about the resonance quantum numbers.

A good test of the production mechanisms might  be the distribution $W(\Theta;\theta,\phi)$ at $\Theta=0$, where the $\Delta$ decay is observed
at forward direction in coincidence with $\eta$ moving in the opposite direction along the beam axis. Then as follows from Eq.\,(\ref{55}) the
matrix $\rho$ becomes diagonal and the $\phi$ dependence in $W(\Theta;\theta,\phi)$ disappears (at $\Theta=0$ the cross section is obviously
invariant under rotation around the $Z$ axis)
\begin{eqnarray}\label{80}
&&W(\Theta=0;\theta,\phi)=\\
&&\phantom{xx}=\frac{3}{4\pi}\,\Big(\rho_{00}(0)\cos^2\theta+\frac{1}{2}(1-\rho_{00}(0))\sin^2\theta \Big)\nonumber\\
&&\phantom{xx}=
\frac{3}{16\pi}\Big(\rho_{00}(0)(1+3\cos^2\theta)+(2-3\rho_{00}(0))\sin^2\theta\Big)\nonumber
\end{eqnarray}
with $\rho_{00}(0)=\rho_{00}(\Theta=0)$.
Using $m=m'=0$ in (\ref{55}) we obtain the following formula for $\rho_{00}(0)$
\begin{equation}\label{85}
\rho_{00}(0)=\frac{2}{3\left(1+ac\right)}
\end{equation}
with
\begin{equation}\label{85a}
c=\left(\frac{C^{J\frac32}_{L0\,\frac32\frac32}}{C^{J\frac12}_{L0\,\frac32\frac12}}\right)^2=3^{2(1+L-J)}\frac{J+\frac32}{J-\frac12}\,,
\quad
a=\left(\frac{A^R_{3/2}}{A^R_{1/2}}\right)^2\,.
\end{equation}
One can readily see from Eqs.~(\ref{80}) to (\ref{85a}) that the first term in Eq.~(\ref{80}), proportional to $1+3\cos^2\theta$, is provided
exclusively by the $\Delta$ helicity $\mu_\Delta=1/2$ whereas the second term with $\sin^2\theta$ is due to $\mu_\Delta=3/2$.

The expressions for $f(\theta)=2\pi\,W(\Theta=0;\theta,\phi)$ are summarized in Table \ref{ta5}. The situation is particularly simple for
$J=1/2$. In this case $f(\theta)$ does not depend on the electromagnetic part and its form is totally fixed by the $\Delta$ spin. It is
identical to the angular distribution of pions through the $\Delta$ decay in single pion photoproduction. For higher resonances having $J\geq
3/2$, also the substates with the helicity $\mu_\Delta=\pm 3/2$ become populated. As a result, the element $\rho_{00}$ and consequently the
shape of $f(\theta)$ depends on the ratio $a=\left(A^R_{1/2}/A^R_{3/2}\right)^2$. In particular, it is convex upwards (downwards) in the whole
region $-1\leq\cos\theta\leq 1$ if $ac<1$ $(>1)$. According to the formulas in Table~\ref{ta5} the angular distribution for $J^\pi=3/2^+$ and
$5/2^-$ is convex upwards for not too low values of $a$. In the other two cases $J^\pi=3/2^-$ and $5/2^+$ we should observe quite a slight
angular dependence with a convex shape downwards for $a<1$ and $a<3/2$, respectively. Taking, e.g., $A_{3/2}/A_{1/2}=0.81$ for $D_{33}(1700)$
from the PDG~\cite{PDG}, we obtain $f(\theta)\approx 3+\cos^2\theta$.

Besides relatively simple formalism (Eqs.~(\ref{80}) to (\ref{85a})) the measurement at $\Theta=0$ has the advantage that at very forward angles
$\Theta$ the overlap between $\pi N$ and $\eta N$ states becomes minimal. It is especially important at low energies, where the restricted phase
space does not allow the particles, e.g., $\eta$ and $\Delta$ to escape the interaction region before the $\Delta$ decays. Therefore, this
method is a possibility to naturally reduce the corrections appearing when the $\Delta$ decay is influenced by the presence of the $\eta$ meson.

As a next step we consider the distribution over the angles $\theta$ and $\phi$ of the pion momentum in the $\pi N$ rest frame. First, we
introduce a new distribution function defined as
\begin{eqnarray}\label{95}
\widetilde{W}(\Omega)&=&
\frac{\pi}{N}\!\int
\,\sum\limits_{m_f\lambda}\left|\,t_{m_f\lambda}(\Theta;\Omega)\right|^2\,\sin\Theta\,d\Theta\,,
\nonumber\\
&&\int \widetilde{W}(\Omega)\,d\Omega=1\,,
\end{eqnarray}
with the normalization constant
\begin{equation}\label{100}
N=\left|\,\alpha^R\right|^2\Big(\left|\,A^R_{1/2}\right|^2+\left|\,A^R_{3/2}\right|^2\Big)\,.
\end{equation}
It is clear that the general structure of $W(\Theta;\Omega)$ (\ref{75}) also holds for $\widetilde{W}(\Omega)$  so that we can immediately write
\begin{eqnarray}\label{102}
&&\widetilde{W}(\Omega)=\sum_{mm'}Y^*_{1m}(\Omega)\tilde{\rho}_{mm'}Y_{1m'}(\Omega)\nonumber\\
&&\phantom{xx}=\frac{3}{4\pi}\,\Big(\tilde{\rho}_{00}\cos^2\theta+\frac12(1-\tilde{\rho}_{00})\sin^2\theta -\sqrt{2}{\cal
R}e\tilde{\rho}_{10}\,\sin2\theta\cos2\phi-\tilde{\rho}_{1-1}\,\sin^2\theta\cos2\phi\Big)\,.
\end{eqnarray}

Due to parity conservation the third term in the brackets proportional to $\sin 2\theta$ should vanish, because it changes sign under the
transformation $\theta\to\pi-\theta,\,\phi\to\phi+\pi$. Therefore $\tilde{\rho}_{10}=0$ and the matrix $\tilde{\rho}$ has only two independent
elements $\rho_0\equiv\tilde{\rho}_{00}$ and $\rho_1\equiv\tilde{\rho}_{1-1}$ and is of the form
\begin{equation}\label{105}
\tilde{\rho}=\left(
\begin{array}{ccc}
  \frac12(1-\rho_0) & 0 & \rho_1 \\
  0 & \rho_0 & 0 \\
  \rho_1 & 0 & \frac12(1-\rho_0)
\end{array}
\right)\,.
\end{equation}
It is instructive to consider the eigenvalues $\alpha$, $\beta$, and $\gamma$  of the matrix $\tilde{\rho}$
\begin{eqnarray}\label{110}
\alpha=\frac12(1-\rho_0)-\rho_1\,,\quad
\beta&=&\frac12(1-\rho_0)+\rho_1\,, \quad \gamma=\rho_0\,,\nonumber\\
\alpha&+&\beta+\gamma=1\,.
\end{eqnarray}
The eigenvalue $\beta$ is proportional to the combination (\ref{70a}) integrated over $\Theta$. From (\ref{110}) and positive definition of the
matrix $\tilde{\rho}$ the following restrictions hold in any reference system
\begin{equation}\label{115}
0\leq\rho_0\leq 1\,,\quad \left|\rho_1\right|\leq\frac12(1-\rho_0)\,.
\end{equation}
These inequalities can be useful in reconstructing the matrix $\tilde{\rho}$ (\ref{105})  from the experimental angular distributions on
$\theta$ and $\phi$.

Using Eq.\,(\ref{20a}) we obtain the following formula for the correlation coefficients in the helicity frame
\begin{equation}\label{120}
\tilde{\rho}^H_{mm'}=\frac{1}{2 N}\sum_{m_f\lambda}\sum_{\mu\mu'} {\cal A}^{\,\mu\mu'}_{m_f\lambda,mm'} c^{J}_{\lambda,\mu\mu'}\,,\quad
\sum_m\tilde{\rho}^H_{mm}=1\,,
\end{equation}
where
\begin{equation}\label{125}
{\cal A}^{\mu\mu'}_{m_f\lambda,mm'}=C^{J\mu}_{L0\,\frac32\mu} C^{J\mu'}_{L0\,\frac32\mu'}C^{\frac32\mu}_{1m\,\frac12m_f}C^{\frac32\mu'}_{1m'\,
\frac12m_f} \left|\,\alpha^RA^R_\lambda\right|^2,
\end{equation}
\begin{eqnarray}\label{130}
&&c^{J}_{\lambda,\mu\mu'}=\frac{2J+1}{2}\int\limits_0^\pi
d^J_{\lambda\mu}(\Theta)\,d^{J}_{\lambda\mu'}(\Theta)\,\sin\Theta\,d\Theta\,, \nonumber\\
&&c^{J}_{\lambda,-\mu'-\mu}=c^{J}_{\lambda,\mu\mu'}\,.
\end{eqnarray}
Substituting $m=m'=0$ and $m=-m'=1$ into (\ref{120}) it is straightforward to obtain the values of $\rho_0$ and $\rho_1$ for an individual
partial wave $R(J^\pi)$. In particular, the parameter $\rho_0$ is of very simple form
\begin{equation}\label{135}
\rho^H_0\equiv\tilde{\rho}^H_{00}=\frac43\left(\frac{2L+1}{2J+1}\right)
\Big(C^{J\frac12}_{L0\,\frac32\frac12}\Big)^2\,,
\end{equation}
so that the diagonal elements of the matrix $\tilde{\rho}$ (\ref{105})  are totally independent of the electromagnetic part and are fixed only
by the spin-parity of the resonance $R$. For $\rho^H_1$ we have
\begin{eqnarray}\label{140}
&&\rho^H_1\equiv\tilde{\rho}^H_{1-1}=\frac{2L+1}{4(1+a)}\sum_{m_f=\pm\frac12} \Big(\sum_{\lambda=\pm\frac12}c^{J}_{\lambda,\mu\mu-2}
+a\sum_{\lambda=\pm\frac32}\,c^{J}_{\lambda,\mu\mu-2}\Big)\nonumber\\
&&\phantom{xx}\times C^{J\mu}_{L0\,\frac32\mu}
C^{J\mu-2}_{L0\,\frac32\mu-2}C^{\frac32\mu}_{11\,\frac12m_f}C^{\frac32\mu-2}_{1-1\,\frac12m_f}\,,
\end{eqnarray}
or using symmetry properties of the Clebsh-Gordan coefficients and the relation  (\ref{130}) for the coefficients $c^{J}_{\lambda,\mu\mu'}$
\begin{equation}\label{145}
\rho^H_1=\frac{2L+1}{1+a}\frac{1}{\sqrt{3}}
\Big(c^{J}_{\frac12,\frac32-\frac12} +a\,c^{J}_{\frac32,\frac32-\frac12}\Big)
\,C^{J\frac32}_{L0\,\frac32\frac32} C^{J-\frac12}_{L0\,\frac32-\frac12}\,.
\end{equation}
Since $c^{\frac12}_{\frac12,\frac32-\frac12}=0$, the resonances with $J^\pi=\frac12^\pm$ do not exhibit $\phi$ dependence. Furthermore,
$c^{\frac52}_{\frac32,\frac32-\frac12}=0$, so that for $J=5/2$ the second term in the brackets, corresponding to the initial helicity state
$\lambda=3/2$, does not contribute. This leads to the fact that the sign of $\rho_1$ for $J^\pi=5/2^\pm$ does not depend on the parameter $a$
and therefore the slope of the corresponding $\phi$-distribution at $\phi\to 0$ (or $\phi\to\pi$) is model independent.

In the canonical system
\begin{equation}\label{150}
\tilde{\rho}^K_{mm'}=\frac{1}{2 N}\sum_{m_f\lambda}\sum_{M_LM_L'}
{\cal B}^{M_LM_L'}_{m_f\lambda,mm'}\, c^{L}_{M_LM_L'}\,,\quad
\sum_m\tilde{\rho}^K_{mm}=1\,,
\end{equation}
where
\begin{equation}\label{155}
{\cal B}^{M_LM_L'}_{m_f\lambda,mm'}=C^{J\lambda}_{LM_L\,\frac32M_\Delta} C^{J\lambda}_{LM_L'\,\frac32M_\Delta'}C^{\frac32M_\Delta}_{1m\,
\frac12m_f}C^{\frac32M_\Delta'}_{1m'\,\frac12m_f} \left|\,\alpha^RA^R_\lambda \right|^2\,,
\end{equation}
\begin{equation}\label{160}
c^{L}_{M_LM_L'}=\frac{2L+1}{2}\int\limits_0^\pi d^L_{M_L0}(\Theta)\,
d^{L}_{M_L'0}(\Theta)\,\sin\Theta\,d\Theta\,.
\end{equation}
The elements $\rho_0^K$ and $\rho_1^K$ are then given by
\begin{eqnarray}\label{165}
\rho^K_0&=&\frac{1}{2N}\sum_{m_f\lambda M_L}{\cal B}^{M_LM_L}_{m_f\lambda,\, 00}\,,
\nonumber\\
\rho^K_1&=&-\frac{1}{4N}\sum_{m_f\lambda}\Big( {\cal B}^{-11}_{m_f\lambda,\, 1-1}+\frac{1}{\sqrt{6}} \sum_{M_L\ne-1} {\cal
B}^{M_LM_L+2}_{m_f\lambda,\, 1-1}\Big)\,.
\end{eqnarray}
Using the last equations one can easily find
\begin{equation}\label{170}
\rho^K_0=\frac{2}{3\,(1+a)}\sum_{m_f=\pm\frac12}\bigg( \Big(C^{J\frac12}_{L\frac12-m_f\,\frac32
m_f}\Big)^2+a\,\Big(C^{J\frac32}_{L\frac32-m_f\,\frac32 m_f}\Big)^2\ \bigg)\,,
\end{equation}
\begin{equation}\label{175}
\rho^K_1=-\frac{1}{\sqrt{3}}\Big( C^{J\frac12}_{L-1\frac32\frac32}C^{J\frac12}_{L1\frac32-\frac12}+
\frac{1}{\sqrt{6}}\,C^{J\frac12}_{L0\frac32\frac12}C^{J\frac12}_{L2\frac32-\frac32}
+\frac{a}{\sqrt{6}}\,C^{J\frac32}_{L0\frac32\frac32}C^{J\frac32}_{L2\frac32-\frac12}\Big)\,.
\end{equation}
It follows from (\ref{175}) that the amplitude $A^R_{3/2}$ can contribute to the  nondiagonal term $\rho^K_1$ only if $L\geq 2$. In our case
this is only $J^\pi=5/2^-$ (see Table\,\ref{ta1}).

The values $\rho^{K/H}_0$ and $\rho^{K/H}_1$ for all six states are given in Table~\ref{ta3}. One can see that the combination $\rho_0-2\rho_1$
does not change if we turn from the $K$- to the $H$-system. This fact becomes trivial if we notice that the equality
$\rho^H_0-2\rho^H_1=\rho^K_0-2\rho^K_1$ immediately follows from the Eq.\,(\ref{70b}) after integration over $\Theta$.

Now let us consider the distribution over $\cos\theta$. Its structure is similar to that in the previously discussed case (\ref{80})
\begin{eqnarray}\label{180}
\widetilde{W}(\theta)&=&\int \widetilde{W}(\theta,\phi)\,d\phi \nonumber\\
&=&\frac{3}{8}\,\Big(\rho_0(1+3\cos^2\theta)+(2-3\rho_0)\sin^2\theta\Big)\,.
\end{eqnarray}
In analogy to $W(\Theta=0;\Omega)$ the first and the second terms in Eq.\,(\ref{180}), evaluated in the $H$-system, are related to the $\Delta$
helicities $|\mu_\Delta|=1/2$ and $|\mu_\Delta|=3/2$, respectively. It is important that in the $H$-system the element $\rho^H_0$ is independent
of $a$ and is totally determined by the quantum numbers of $R$ (see Table~\ref{ta3}). Furthermore, resonances with different angular momentum
and parity contribute incoherently to $\rho^H$. Therefore, the function $\widetilde{W}(\theta)$ calculated in the helicity frame is especially
effective as a tool to identify $R$.

As for the $\phi$ dependence of the cross section, its structure follows from the general expression (\ref{102})
\begin{equation}\label{185}
\widetilde{W}(\phi)=\int \widetilde{W}(\theta,\phi)\sin\theta\,
d\theta=\frac{1}{2\pi}\big(1-2\rho_1\cos 2\phi\big)\,,
\end{equation}
where $\rho_1$ in $H$- and $K$-systems are given by the Eqs.\,(\ref{145}) and (\ref{175}).  As already pointed out, the $\phi$ dependence for
$\frac12^\pm$ is trivial in the helicity system. Furthermore, for the highest resonances with $J^\pi=\frac52^\pm$, although the amplitude of the
oscillations of $\widetilde{W}$ depends on the parameter $a$ the character of its convexity in the region $0\leq\phi\leq\pi$ is independent of
$a$. In the $K$-system, the $3/2$-helicity amplitude contributes to $\rho_1$ only for $J^\pi=5/2^-$. In the whole, as one can see from
Table~\ref{ta3}, the sign of $\rho_1$ is fixed only by the spin-parity of the resonance and the character of the $\phi$ distribution is model
independent.

\begin{table*}
\renewcommand{\arraystretch}{2.0}
\caption{Coefficients $\rho_0\equiv\tilde{\rho}_{00}$ and $\rho_1\equiv\tilde{\rho}_{1-1}$ in Eq.(\protect\ref{102}) (see also
(\ref{180}) and (\ref{185}))
calculated in the $H$- and the $K$-system.} \label{ta3}
\begin{center}
\begin{tabular*}{16cm}{@{\hspace{0.7cm}}c@{\hspace{0.7cm}}|@{\hspace{0.8cm}}c@{\hspace{0.8cm}}c@{\hspace{0.8cm}}c@{\hspace{0.8cm}}
c@{\hspace{0.8cm}}c@{\hspace{0.8cm}}c@{\hspace{0.8cm}}} \hline\noalign{\smallskip}
$J^\pi(L_{2T2J})$ & $\frac12^-(S_{31})$ & $\frac12^+(P_{31})$& $\frac32^-(D_{33})$ & $\frac32^+(P_{33})$ & $\frac52^-(D_{35})$ & $\frac52^+(F_{35})$\\
\noalign{\smallskip}\hline
\noalign{\smallskip}\hline\noalign{\smallskip}
$\rho_0^H$ & $\frac23$ & $\frac23$  & $\frac13$ & $\frac{1}{15}$ & $\frac{2}{21}$ & $\frac25$ \\
\noalign{\smallskip}\hline\noalign{\smallskip}
$\rho_1^H$ & 0 & 0 & $-\frac16\frac{1-a}{1+a}$ & $\frac{1}{10}\frac{1-a}{1+a}$ & $\frac17\frac{1}{1+a}$ & $-\frac{1}{5}\frac{1}{1+a}$ \\
\noalign{\smallskip}\hline
\noalign{\smallskip}\hline
$\rho_0^K$ & $\frac13$ & $\frac13$  & $\frac23\frac{1}{1+a}$ & $\frac{2}{15}\frac{3+2a}{1+a}$ & $\frac{1}{105}\frac{31+34a}{1+a}$ & $\frac{1}{5}\frac{3+2a}{1+a}$ \\
\noalign{\smallskip}\hline\noalign{\smallskip}
$\rho_1^K$ & $-\frac16$ & $-\frac16$ & $0$ & $\frac{4}{15}\frac{1}{1+a}$  & $\frac{1}{70}\frac{17+8a}{1+a}$ & $-\frac{1}{10}\frac{1}{1+a}$ \\
\noalign{\smallskip}\hline
\end{tabular*}
\end{center}
\end{table*}

\section{Isobar model for $\gamma N\to \pi^0\eta N$. Discussion of the results.}\label{results}

The expressions presented in the previous section relate to the ideal situation where the amplitude for $\gamma N\to\pi^0\eta N$ is dominated by
the single diagram, Fig.~\ref{fig1}(g). The natural question arises: what is the influence of the $\pi N^*$ channel? Inserting the corresponding
diagram, Fig.~\ref{fig1}(h), into the formulas above will certainly make them much lengthier and less symmetric. Therefore we consider the
problem numerically and discuss in this section the influence of the $\pi N^*$ production on the results predicted by the formalism of
Sect.\,\ref{model}. For further study we need a model describing photoexcitation of the state $R(J^\pi)$ and its decay according to the schemes
(a) and (b) in Eq.\,(\ref{15}). Here we adopt a typical isobar model along the line used for double pion production (see, e.g.,
\cite{Oset,Ochi,FiAr2pi}).

The amplitude used in the calculation is a sum of the eight terms corresponding to the diagrams in Fig.~\ref{fig1}
\begin{equation}\label{186}
T=T^{(a\mbox{-}f)}+T^{(g)}+T^{(h)}.
\end{equation}
The first six graphs (a-f) form the background. We neglect the diagrams with $\eta NN$ coupling due to its weakness. Therefore, the main model
parameters are the partial widths of $N^*$. We use 45\,$\%$ for both $\eta N$ and $\pi N$ modes and 10\,$\%$ for that of $\pi\pi N$. The total
width is equal to $\Gamma_{N^*}=150$ MeV. As already mentioned, the background mechanisms provide only a small fraction of the observed cross
section for $\gamma p\to\eta\pi^0 p$, and this fact is considered as a key indication that the reaction mainly proceeds through resonance
excitation. The corresponding amplitudes, depicted in Fig.~\ref{fig1}(g-h), read
\begin{equation}\label{187a}
T^{(g)}=T_{\gamma N\to\eta\Delta}G_\Delta F_{\Delta\to\pi N}\,,
\end{equation}
\begin{equation}\label{187b}
T^{(h)}=T_{\gamma N\to\pi N^*}G_{N^*} F_{N^*\to\eta N}\,,
\end{equation}
with $G_\Delta$ and $G_{N^*}$ standing for the $\Delta$ and $N^*$ propagators.

For each resonance $R(J^\pi)$ we used a simple nonrelativistic Breit-Wigner ansatz with an energy-dependent width $\Gamma_R(W)$
\begin{eqnarray}\label{188}
&&\langle \lambda|T_{\gamma N\to \eta\Delta}|\vec{q},M_\Delta\rangle= A^R_\lambda\,G_R\,\langle J\lambda\,|F_{R\to
\eta\Delta}|\vec{q},M_\Delta\rangle\,,
\nonumber\\
&&\nonumber\\
&&\langle \lambda|T_{\gamma N\to \pi N^*}|\vec{q},M_{N^*}\rangle= A^R_\lambda\,G_R\,\langle J\lambda\,|F_{R\to\pi
N^*}|\vec{q},M_{N^*}\rangle\,,
\nonumber\\
&&G_R=\left(W-M_R+\frac{i}{2}\Gamma_R(W)\right)^{-1},\quad \lambda=1/2,\,3/2\,.
\end{eqnarray}
The vertices $F_{R\to x}$\,$(x\in\{\eta\Delta,\,\pi N^*\})$ in Eq.\,(\ref{188}) were taken in the phenomenological form
\begin{eqnarray}
\label{13b} &&\langle JM|F_{R\to\eta\Delta}|\,\vec{q},M_\Delta\rangle=f_{R\eta\Delta}\frac{q^L}{m_\pi^L}\,
C^{JM}_{LM_L\,\frac32M_\Delta}Y^*_{LM_L}(\hat{q})\,,\quad L=L(\eta\Delta)\,,\\
\label{13d} &&\langle JM|F_{R\to\pi N^*}|\,\vec{q},M_{N^*}\rangle= f_{R\pi N^*}\frac{q^L}{m_\pi^L}\,
C^{JM}_{LM_L\,\frac12M_{N^*}}Y^*_{LM_L}(\hat{q}),\quad L=L(\pi N^*)\,,
\end{eqnarray}
where $L(\eta\Delta)$ and $L(\pi N^*)$ are given in Table~\ref{ta1}. As already noted, in (\ref{13b}) we assume only the lower of two possible
values of the angular momentum $L(\eta\Delta)$. In both channels (\ref{13b}) and (\ref{13d}) the finite width of the $\Delta$ and $N^*$ isobars
was taken into account, what is important for the low energies considered here. Significant contribution to the width of each resonance is
assumed to come from the $\pi\pi N$ mode (see Eq.\,(\ref{195})). The corresponding energy dependence was taken in a simple form
\begin{equation}\label{189}
\Gamma_{\pi\pi N}(W)\sim (W-M_N-2m_\pi)\,\Theta(W-M_N-2m_\pi).
\end{equation}
Since our calculation relates to the region of low kinetic energies we use the nonrelativistic formalism for $\Delta$ and $N^*$ states.
Therefore, we do not touch upon such a complication as off-shell ambiguity in the $F_{\Delta\to\pi N}$ vertex in Eq.\,(\ref{187a})
appearing in the relativistic treatment of the spin 3/2 field (see, e.g., Ref.~\cite{Mukh}).

Taking the $R\to\pi N^*$ transition in the phenomenological form (\ref{13d}) we have essentially simplified the problem in comparison to
Ref.~\cite{Doring} where the $D_{33}\to\pi N^*$ decay is calculated microscopically. Within the approach of Ref.~\cite{Doring} the process
$\gamma N\to D_{33}\to\eta\Delta$ is treated as a driving mechanism producing the $\pi^0\eta N$ state already at tree level. The coupling
constant $f_{R\eta\Delta}$ entering the vertex $R\to\eta\Delta$ is taken from the analysis of Ref.~\cite{Sarkar}. Then the production of the
$\pi N^*$ state proceeds as a series of interactions $D_{33}\to\eta\Delta^+\to\pi^0\eta p\to\pi^0N^*$, taken up to the first order in the
corresponding two-body scattering matrices. In our case, the constant $f_{R\pi N^*}$ in (\ref{13d}) is real and its absolute value is fixed by
the unitarity condition which for the Breit-Wigner resonance reads
\begin{equation}\label{190}
\Gamma_{\pi N^*}=\Gamma_R-\Gamma_{\pi N}-\Gamma_{\pi\pi N}-\Gamma_{\eta\Delta}\,.
\end{equation}
The model should be reasonably good at least close to the resonance position $W\approx M_R$. With distance from this point the energy dependence
dictated by the Breit-Wigner ansatz~(\ref{188}) could differ from the one obtained within the microscopic approach. Clearly, the most
unambiguous treatment would be a three-body calculation including all coupled channels $\pi\pi N$, $\pi N$ and $\pi \eta N$.

Close to the $\pi\eta$ production threshold, the functions $\Gamma_{\eta\Delta}(W)$ and $\Gamma_{\pi N^*}(W)$ are mainly determined by the
centrifugal barrier effect resulting in $\Gamma_x\sim q_x^{2L(x)+1}$, where $x\in\{\eta \Delta, \pi N^*\}$ and the orbital momenta $L(x)$ are
collected in Table~\ref{ta1}. We can expect that in the low energy region the relative fraction of the $\pi N^*$ channel is important in the
$\frac12^\pm$ channels since their decaying into $\eta\Delta$ requires higher values of $L(\eta\Delta)$. As already noted, the $\frac32^-$ state
might be a well candidate to explain the rapid rise of the cross section in the threshold region. Another state producing $s$-waves in the
$\pi\eta N$ system is $\frac12^+$. Any appreciable amount of other states is less likely since their decay into $\pi\eta N$  at low energies is
suppressed by the centrifugal barrier.

\begin{figure}
\begin{center}
\resizebox{0.35\textwidth}{!}{%
\includegraphics{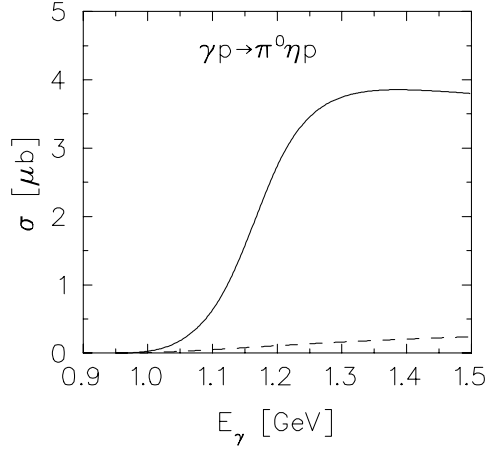}}
\caption{Total cross section for $\gamma p\to\pi^0\eta p$. The
dashed curve is the background contribution (diagrams (a) to (f) in
Fig.\,\protect\ref{fig1}).}
\label{figTot}
\end{center}
\end{figure}

In Fig.\,\ref{figTot} we show an example of the total cross section calculated with $J^\pi=\frac32^-$. The calculation demonstrates a strong
dominance of the resonance mechanism over the background terms (dashed curve). This result agrees with that of Ref.\,\cite{Doring}.

For the helicity amplitudes of $D_{33}(1700)$ we used average values~\cite{PDG}
\begin{equation}\label{196}
A_{1/2}=0.104\ \mbox{GeV}^{-1/2}\,,\quad A_{3/2}=0.085\ \mbox{GeV}^{-1/2}\,,
\end{equation}
and for the mass and widths
\begin{eqnarray}\label{195}
&&M_R=1.72\,\mbox{GeV}\,,\quad \Gamma_R=300\,\mbox{MeV}\,,
\quad \frac{\Gamma_{\pi N}}{\Gamma_R}=20\,\%\,,\nonumber\\
&&\frac{\Gamma_{\pi \pi
N}}{\Gamma_R}=73\,\%\,,\quad \frac{\Gamma_{\eta\Delta}}{\Gamma_R}=5\,\%\,,
\quad \frac{\Gamma_{\pi N^*}}{\Gamma_R}=2\,\%\,,
\end{eqnarray}
where all $\Gamma$'s are taken at $W=M_R$. The last three values were chosen simply by adjusting the resulting total cross section for $\gamma
p\to\pi^0\eta p$ to the data of Ref.~\cite{PiEtaDat}. It is remarkable that fitting the cross section may require quite a small strength of
$D_{33}(1700)$ decay into the $\eta\Delta$ and $\pi N^*$ channels. In order to get a feeling of possible variation of $\Gamma_{\eta\Delta}$ and
$\Gamma_{\pi N^*}$ one can use the formula for the total cross section at the resonance position
\begin{equation}\label{198}
\sigma(M_R)= C_T(2J+1)\frac{\pi}{k^2}\frac{\Gamma_{\gamma N}\Gamma_{\pi\eta N}}{\Gamma_R^2}\,,
\end{equation}
where all energy dependent quantities are calculated at $W=M_R$. The coefficient $C_T$ takes into account the isospin structure and
$\Gamma_{\gamma N}$ is the radiation decay width. Taking for $D_{33}(1700)$: $C_T$=4/9, $M_R=1.7$ GeV, $\Gamma_R=300$ MeV, and $\Gamma_{\gamma
N}/\Gamma_R=0.19\,\%$ from \cite{PDG} we obtain
\begin{equation}\label{198a}
\frac{\Gamma_{\pi\eta N}}{\Gamma_R}\approx 8.5\cdot 10^{-2}\sigma(M_R)\,.
\end{equation}
Since around the total energy $W=1.7$ GeV the cross section has a strong rise the ratio (\ref{198a}) is very sensitive to $M_R$. For instance,
for $M_R=1.72$ GeV we will have $\sigma(M_R)\approx 0.8$\,$\mu$b and $\Gamma_{\pi\eta N}=7\%\,\Gamma_R$. The maximum value $\sigma\approx
4$\,$\mu$b \cite{PiEtaDat} gives about 34$\%$ for the total $\pi\eta N$ width. Then the inequality
\begin{equation}
\Gamma_{\eta\Delta}+\Gamma_{\pi N^*}\leq\Gamma_{\pi\eta N}=0.34\,\Gamma_R
\end{equation}
holding for the constructive interference between $\eta\Delta$  and $\pi N^*$ configurations gives an upper limit for the sum of the partial
decay widths in $\eta\Delta$ and $\pi N^*$.

\begin{figure}
\begin{center}
\resizebox{0.47\textwidth}{!}{%
\includegraphics{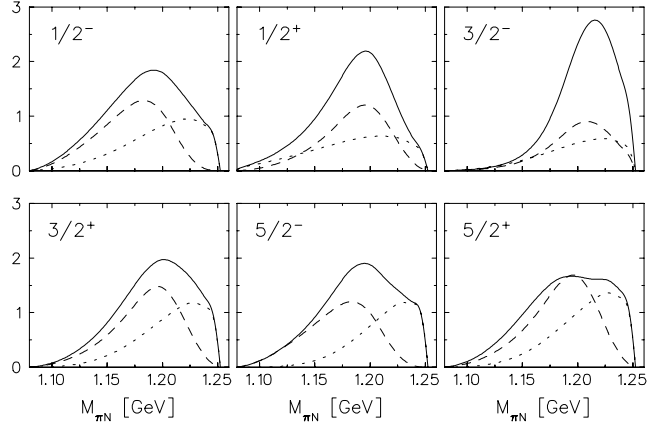}}
\caption{Distribution of the $\pi N$ invariant  mass calculated at the
total $\gamma N$ c.m.\ energy $W=1.8$ GeV (corresponds to the photon
lab energy about 1.255 GeV). The dashed and the dotted lines show the
contribution from the $\eta\Delta$ and $\pi N^*$ production. The
solid line is the coherent sum of both channels. The results are
presented in arbitrary units.}
\label{figwPiN}
\end{center}
\end{figure}

\begin{figure}
\begin{center}
\resizebox{0.47\textwidth}{!}{%
\includegraphics{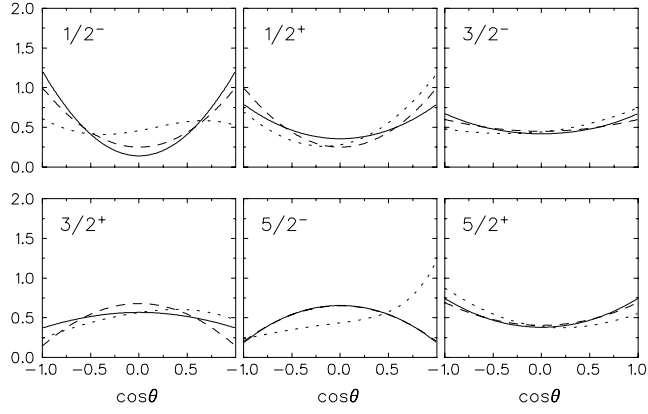}}
\caption{Angular distributions $f(\theta)=W(\Theta=0;\theta,\phi)$
of $\pi$ mesons in the $\pi N$ c.m.\ system when the angle $\Theta$
is fixed to $\Theta=0$. The dashed curve contains only the
contribution of the driving $\eta\Delta$ term in $R\to\pi\eta N$
decay (the channel $(a)$ in Eq.~(\protect\ref{15})). In the dotted curve also
the $\pi N^*$ channel ($(b)$ in Eq.~(\protect\ref{15})) is taken into
account. The solid curve represents the symmetrized function
$f_S(\theta)$ (see Eq.\,(\protect\ref{210})) where in addition the
invariant $\pi N$ energy is restricted to the region $M_{\pi N}<
1.13$~GeV. The total $\gamma N$ c.m.\ energy is $W=1.8$ GeV.}
\label{figQ0}
\end{center}
\end{figure}

\begin{figure}
\begin{center}
\resizebox{0.47\textwidth}{!}{%
\includegraphics{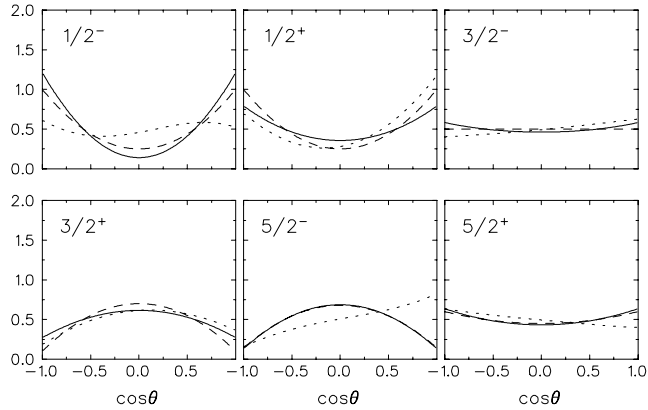}}
\caption{Distribution $\widetilde{W}(\theta)$ (Eq.\,(\ref{180}))
over the polar pion angle in the $\pi N$ rest frame calculated in
the helicity system at $W=1.8$ GeV. Notation of the curves as in
Fig.\,\protect\ref{figQ0}.}
\label{figxH}
\end{center}
\end{figure}

\begin{figure}
\begin{center}
\resizebox{0.47\textwidth}{!}{%
\includegraphics{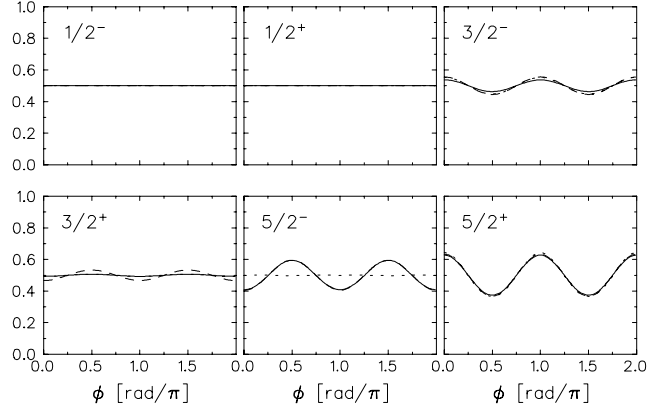}}
\caption{Distribution $\pi\widetilde{W}(\phi)$ (Eq.\,(\ref{185}))
over the azimuthal pion angle in the $\pi N$ rest frame calculated
in the helicity system at $W=1.8$ GeV. Notation of the curves as in
Fig.\,\protect\ref{figQ0}.}
\label{figfiH}
\end{center}
\end{figure}

\begin{figure}
\begin{center}
\resizebox{0.47\textwidth}{!}{%
\includegraphics{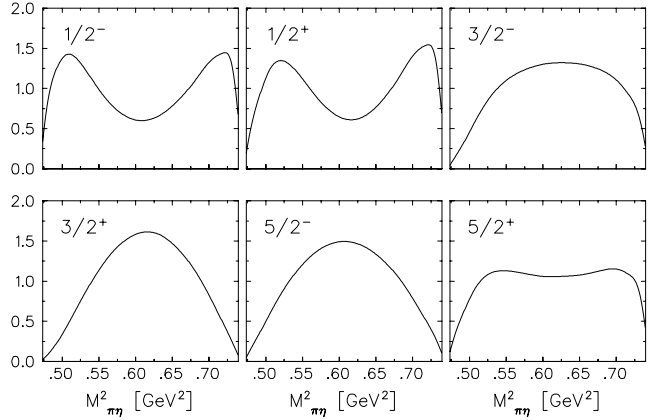}}
\caption{$\pi^0\eta$ invariant mass spectrum at $W=1.8$ GeV  for
different transitions listed in Table~\protect\ref{ta1} in arbitrary
units. The results are obtained without taking the $\pi N^*$ channel
into account.}
\label{fig4}
\end{center}
\end{figure}

In the following we discuss a general case in which a resonance $R(J^\pi)$ produces $\pi\eta N$ according to the two schemes in Eq.\,(\ref{15})
and the background is totally neglected. For each resonance $R(J^\pi)$ we use the same parameters
\begin{eqnarray}\label{197}
&&M_R=1.8\,\mbox{GeV}\,,\quad \Gamma_R=300\,\mbox{MeV}\,,\quad \frac{\Gamma_{\pi N}}{\Gamma_R}=20\%\,,\nonumber\\
&&\frac{\Gamma_{\pi \pi N}}{\Gamma_R}=60\%\,,\quad
\frac{\Gamma_{\eta\Delta}}{\Gamma_R}=20\%\,,\quad \frac{\Gamma_{\pi N^*}}{\Gamma_R}=10\%\,.
\end{eqnarray}
For the ratio $a$~(\ref{35a}) we take
\begin{equation}\label{200}
a\equiv\left(\frac{A_{3/2}}{A_{1/2}}\right)^2=0.67\,.
\end{equation}
Firstly, we show in Fig.\,\ref{figwPiN} the $\pi N$  invariant mass spectrum where the contributions of the final state configurations
$\eta\Delta$ and $\pi N^*$ (schemes (a) and (b) in Eq.\,(\ref{15})) are separately presented. It is interesting that the overlap of these states
essentially differs in different partial waves. It is quite large in $J^\pi=\frac32^-$ and $\frac52^+$ and less essential in other waves.
Clearly, the character of the interference depends on the particular values of the orbital momenta $L(\eta\Delta)$ and $L(\pi N^*)$ as well as
on the relative sign of the $R\eta\Delta$ and $R\pi N^*$ coupling constants.

Of special importance for us is the contribution of the $\pi N^*$ state in the region of the low invariant masses $M_{\pi N}$. As we can see in
Fig.\,\ref{figwPiN} it is important in $\frac12^+$ and $\frac32^-$ states and can be neglected in other waves if sufficiently low values of
$M_{\pi N}$ are considered. Therefore, to isolate the dynamics related to the scheme (a) in Eq.\,(\ref{15}) we only need to exclude the region
with $M_{\pi N}$ larger than a certain value $M_{\pi N}^0$ depending on the overall reaction energy $W$. Then it is reasonable to assume that
the $\Delta$ life time is sufficient to escape interaction with the $\eta$ meson. In the waves with $J^\pi=\frac12^+$ and $\frac32^-$  the
situation is more complicated at least at the energy $W=1.8$~GeV considered here. Clearly the overlap between $\eta \Delta$ and $\pi N^*$ should
decrease with increasing $W$.

In the series of figures \ref{figQ0} to \ref{figfiH}  we present examples of angular distributions for $\pi^0\eta$ photoproduction by
unpolarized photons on an unpolarized nucleon. The calculation is performed at $W=1.8$~GeV. In all figures the dashed curve represent the
distribution, in which only the channel (a) in Eq.\,(\ref{15}) is taken into account. Their form is described by the analytic expressions
obtained in Sect.\,\ref{model}. Addition of the scheme (b) gives the dotted curve and the solid curve is obtained after cutting off the
kinematical region with $M_{\pi N}\geq M_{\pi N}^0=1.13$~GeV. As expected, the interference between the channels (a) and (b) tends to distort
simple angular dependence, obtained if only the scheme (a) is used. After eliminating the energy region in which both mechanisms strongly
overlap, we bring the calculation back to qualitative agreement with the results shown by the dashed curves. This effect is observed in all
waves except for $J^\pi=\frac12^+$ and $\frac32^-$ in accord with our notion about strong overlap of the $\eta\Delta$ and $\pi N^*$
configurations in these states. In the whole, using the above procedure the qualitative features of the resulting angular distributions are in
agreement with the simplified calculations, in which the $\pi N^*$ channel is neglected.

The solid curve in Fig.~\ref{figQ0}
contains symmetrization with respect to $\cos\theta=0$, i.e.
\begin{equation}\label{210}
f(\theta)\to f_S(\theta)=
\frac12\Big(f(\theta)+f(\pi-\theta)\Big)\,.
\end{equation}
In general, after addition of the channel $\pi N^*$  the resulting angular dependence yields certain forward-backward asymmetry, which can make
the analysis more complicated. This effect is removed after the artificial symmetrization (\ref{210}).

The $\theta$-distribution in Fig.\,\ref{figxH} is quite similar to the one in Fig.\,\ref{figQ0}. As already noted the convex up and down form of
$\widetilde{W}(\theta)$ related to $\sin^2\theta$ and $(1+3\cos^2\theta)$ terms in Eq.\,(\ref{80}) is provided by the $\Delta$ helicities
$|\mu_\Delta|=3/2$ and $|\mu_\Delta|=1/2$. Flat distribution for $J^\pi=\frac32^-$ is due to the fact that in this instance (after integration
over $\cos\Theta$) we have an even mixture of both helicity states so that the sum of $\frac13(1+3\cos^2\theta)$ and $\sin^2\theta$ gives a
constant value. In other words, in the $H$-system the longitudinal $\Delta$ polarization (averaged over the region $-1\leq\cos\Theta\leq 1$) is
zero if $J^\pi=3/2^-$. In the case of a small admixture of the $\pi N^*$ background under the $\Delta$ peak we will have an interference term
proportional to $\cos\theta$ caused by different parities of $N^*$ and $\Delta$. This effect is clearly seen in Fig.\,\ref{figxH} (dotted line
in the panel for $J^\pi=\frac32^-$). In other cases addition of the $\pi N^*$ channel leads to more complicated form of the function
$\widetilde{W}(\theta)$.

Knowledge of the angular distribution $\widetilde{W}(\theta)$  in the helicity system is of key importance for understanding the spectrum of the
$\pi\eta$ pairs. Indeed, for each mass $M_{\pi N}$ the value of $\cos\theta$ ($\theta\equiv\theta^*_\pi$) is determined by $M_{\pi\eta}$
\cite{Buk} according to
\begin{equation}\label{212}
\cos\theta=\frac{1}{4WM_{\pi N}q^*_{\pi}q_\eta}\Big[(M_{\pi N}^2-M_N^2+m_\pi^2)(W^2-M_{\pi N}^2-m_\eta^2)-2M_{\pi
N}^2(M_{\pi\eta}^2-m_\pi^2-m_\eta^2)\Big]\,,
\end{equation}
Therefore, the knowledge of $\widetilde{W}(\cos\theta)$ at  fixed $M_{\pi N}$ immediately provides the Dalitz plot distribution\\
$d^2\sigma/dM_{\pi N}dM_{\pi\eta}$, and for the $\pi\eta$ spectrum we have
\begin{equation}\label{214}
\frac{d\sigma}{dM_{\pi\eta}^2}=\int \frac{d^2\sigma}{dM_{\pi N}d\cos\theta}\,
\frac{M_{\pi N}}{2Wq^*_\pi q_\eta}\,dM_{\pi N}\,.
\end{equation}
In other words, if there are no nearby resonances in the $\pi \eta$ system (as in our case), the structure of the Dalitz plot $(M_{\pi
N},M_{\pi \eta})$ is totally determined by the quantum numbers of the resonance $R$ related to its decay into the $\eta\Delta$ and the $\pi N^*$
channels. In Fig.\,\ref{fig4} we present the spectrum $d\sigma/dM^2_{\pi N}$ given by different states $R(J^\pi)$. As we can see, apart from the
boundary of the allowed kinematical region where $d\sigma/dM^2_{\pi\eta}\to 0$ the spectrum qualitatively reproduces the shape of the angular
distribution in the helicity system (dashed curve in Fig.\,\ref{figxH}). Thus the $\pi\eta$ mass distribution should be sensitive to the quantum
numbers of the resonance $R$. Again for $J^\pi=1/2^\pm$ the spectrum, having a visible minimum in the middle part is independent of the
electromagnetic properties of the resonance. For other resonances it depends on the parameter $a=(A^R_{1/2}/A^R_{3/2})^2$.

As for the invariant mass distribution in other two-body subsystems, the corresponding measurements can hardly give useful information. As an
example, we can take $d\sigma/dM_{\pi N}$ shown in Fig.\,\ref{figwPiN}. First of all, the general structure is quite insensitive to the choice
of $R(J^\pi)$. If we change from one resonance to another, in the main only the position of the maximum is shifted. Obviously, this shift is
explained by the barrier effects. Namely, since $d\sigma/dM_{\pi N}\sim d\sigma/dq_\eta$, for low values of $q_\eta$ the spectrum is
proportional to $q_\eta^{2L(\eta\Delta)+1}$, where $L(\eta\Delta)$ is the angular momentum of the decay $R\to\eta\Delta$ (see Table \ref{ta1}).
With increasing $L(\eta\Delta)$ the centrifugal barrier factor tends to suppress the cross section at low $q_\eta$ (large $M_{\pi N}$),
resulting in shifting the maximum to higher values of $q_\eta$ (lower $M_{\pi N}$). This trivial effect is what we mainly observe in
Fig.\,\ref{figwPiN}. Furthermore, at higher energies the shape of the spectrum around $M_{\pi N}=M_\Delta$ will be governed by the $\pi N$
energy distribution in the $\Delta$ region, so that the values of $d\sigma/dM_{\pi N}$ is mainly determined by the form of the $\Delta$ peak. In
this connection, investigation of $d\sigma/dM_{\pi N}$ is not of any use to get additional information on the reaction mechanism.

\section{Conclusion}

We have discussed some details of a phenomenological analysis of $\gamma N\to\pi^0\eta N$ aimed at identifying the dominant mechanisms of this
reaction. This analysis assumes that the part of the amplitude corresponding to a given resonance $R$ is sufficiently large, so that other
partial waves and the background can be neglected. This assumption seems to be justified by direct calculation of the most important Born
diagrams (see Fig.\,\ref{fig1}). Our results show that even in the absence of the polarization data some interesting properties of the reaction
can be found. Our main focus is on the angular dependence as a test of the mechanism responsible for $\pi^0\eta$ photoproduction. Here we
summarize the most important qualitative features of different types of the angular distributions considered in the main part of the paper.

1) The distribution $W(\Theta)$ over the polar angle of the $\pi N$ system in the $\gamma N$ c.m.\ frame (Table\,\ref{ta5}). In the simplest
case of $J^\pi=\frac12^\pm$ we have $W(\Theta)=$ const. Generally, the distribution on $\Theta$ depends on the parameter
$a=(A_{3/2}/A_{1/2})^2$, which hampers the analysis if the electromagnetic properties of the resonance are poorly known. Indirect clues to the
spin-parity of the resonance can be gained from the complexity of $W(\Theta)$. For a state with $\eta\Delta$ angular momentum $L$ the complexity
of the $W(\Theta)$ is fixed to $2L$ for $J\geq \frac32$.

2) The distribution over $\cos\theta$ at fixed $\Theta=0$ (Fig.\,\ref{figQ0}). If we select $\eta$ mesons produced opposite in the direction to
the photon beam, a simple formula (\ref{85}) can be obtained. In the general case the reaction mechanism is such that the intermediate $\Delta$
states with both helicities $|\mu_\Delta|=1/2$ and $3/2$ are populated. As a result, the decay angular distribution of $\Delta$ differs from the
simple form $1+3\cos^2\theta$, peculiar for single pion photoproduction. The weight of each $\mu_\Delta$ configuration depends on $R(J^\pi)$ so
that the method should allow the quantum numbers of $R$ to be extracted from the measurements. As in the $W(\Theta)$ case, the particular form
of the distribution in the states with $J^\pi=\frac32^+$ and $\frac52^\pm$ depends on the value of the parameter $a$, Eq.\,(\ref{35a}). However,
in a wide range of $a$, $a>\frac19$ for $J^\pi=\frac32^+$ and $a>\frac16$ for $J^\pi=\frac52^-$, the corresponding distributions are not very
sensitive to $a$.

3) The distributions $\widetilde{W}(\theta)$ and $\widetilde{W}(\phi)$ over the polar and azimuthal pion angles in the $\pi N$ rest frame. For
convenience, we express the functions $\widetilde{W}(\theta)$ and $\widetilde{W}(\phi)$ in terms of the $\Delta$ decay correlation coefficients
$\rho_{mm'}$. Their values can be determined by fitting the analytic expressions for $\widetilde{W}$ to the experimental data.

For $\widetilde{W}(\theta)$ the helicity system (Fig.~\ref{figxH}) seems to be more useful for the partial wave analysis since the decay
correlation coefficients $\rho_{mm'}^H$ and the corresponding angular distributions are independent of $a$. Furthermore, there is no
interference between partial waves with different spin-parity $J^\pi$, so that resonances contribute incoherently. This is especially important
in the situation of strongly overlapping states, which is quite typical in the second and the third resonance region. However, there is a
sensitivity of the angular distributions to even small admixtures of the $\pi N^*$ channel. Furthermore, in this case there are qualitative
difficulties in distinguishing between the states belonging to the two groups $J^\pi=1/2^\pm,5/2^+$ and $J^\pi=3/2^\pm,5/2^-$.

The distribution $\widetilde{W}(\phi)$ in the $H$-system for $J^\pi=\frac32^\pm$ seems to be sensitive to the parameter $a$, Eq.\,(\ref{35a}).
This shortcoming is however partially avoided in the canonical frame where the sign of the $\phi$-dependent term does not depend on $a$.

It has also been shown that if the correlation coefficients $\tilde{\rho}^H_{00}$ (\ref{135}) are fitted to the data, the spectrum
$d\sigma/dM_{\pi N}$ does not provide additional information, since the Dalitz plot is immediately obtained if the distribution over
$\cos\theta$ in the $H$-system is known. This result is a trivial consequence of a linear relation between the cosine of the pion decay angle
and $M_{\pi\eta}^2$ (see Eq.\,(\ref{212})).

In general, our results demonstrate that different assumptions about the spin-parity of the dominating partial wave lead to different
predictions regarding angular distributions, so that each state $R$ shows its own signature. Model independence of some of these signals as
predicted by the present analysis is the major motivation for proposing our method. Additional important information can be obtained from the
polarization experiments.

Finally, we would like to note, that the presented calculations are related to the case when most of the $\pi^0\eta N$ configurations are
produced through the $R\to\eta \Delta$ decay. On the other hand, rather low percentage of $\eta\Delta$ and $\pi N^*$ channels in
Eq.\,(\ref{195}) indicates that the the measured cross section \cite{PiEtaDat} may be accounted for with quite a small fraction of $\eta\Delta$
and $\pi N^*$ channels in the total resonance width. This observation might be a reason to doubt the necessity of introducing direct
$R\eta\Delta$ coupling to explain $\pi^0\eta$ production rate. The needed strength can be provided by the $R\to\pi\Delta$ and $R\to\rho N$
decays followed by $\pi N\to\eta N$ rescattering. In this case we do not need to restrict ourself to the resonance states with the isospin
$T=3/2$. Clearly, another assumption about the reaction mechanism, as to the one in which the $\pi N^*$ dominates, will lead to angular
distributions different from those presented here.

\section*{Acknowledgment}
The work was supported by the Deutsche Forschungsgemeinschaft (SFB
443) and by the RF Presidential Grant (No MD-2772.2007.2).

\end{document}